# Ballistic transport in induced one-dimensional hole systems

O. Klochan[a], W.R. Clarke, R. Danneau, A.P. Micolich, L.H. Ho and A.R. Hamilton[b]

*School of Physics, University of New South Wales, Sydney NSW 2052, Australia.*

K. Muraki and Y. Hirayama

*NTT Basic Research Laboratory, NTT Corporation, 3-1 Morinosato Wakamiya, Atsugi,*

*Kanagawa 243-0198, Japan*

We have fabricated and studied a ballistic one-dimensional p-type quantum wire using an *undoped* AlGaAs/GaAs heterostructure. The absence of modulation doping eliminates remote ionized impurity scattering and allows high mobilities to be achieved over a wide range of hole densities, and in particular, at very low densities where carrier-carrier interactions are strongest. The device exhibits clear quantized conductance plateaus with highly stable gate characteristics. These devices provide opportunities for studying spin-orbit coupling and interaction effects in mesoscopic hole systems in the strong interaction regime where $r_s > 10$.

---

[a] Author to whom correspondence should be addressed; Electronic mail: klochan@phys.unsw.edu.au

[b] Electronic mail: Alex.Hamilton@unsw.edu.au

High mobility two-dimensional (2D) hole systems in GaAs/AlGaAs heterostructures are of great interest. They enable studies of spin-orbit coupling,[1] due to their p-like valence band, and strong interaction effects,[2] due to their high effective mass $m^*$. Additionally, studies of hole systems could generate further insight into many-body effects such as the anomalous plateau at $0.7 \times 2e^2/h$ in one-dimensional (1D) systems (also known as the 0.7-structure),[3] and lead to advances in spintronic devices. However, difficulties in fabrication, combined with the poor stability of their electrical characteristics, has limited experimental studies of lower-dimensional hole devices until now.[4-7] Only recently have devices using modulation doped AlGaAs/GaAs heterostructures been sufficiently stable to observe the 0.7-structure in 1D wires and single hole transport in quantum dots. The former was achieved using metal gates on a (311)A-oriented AlGaAs/GaAs double quantum well structure[8] and the latter by local anodic oxidation lithography on C-doped (100)-oriented AlGaAs/GaAs heterostructures.[9]

A common feature of all 1D hole devices studied so far is the presence of a modulation doping layer in the heterostructure. In modulation-doped devices, the dominant scattering mechanism at low carrier density is Coulomb interactions with the ionized dopants. This remote ionized dopant scattering has two key effects: (a) it causes non-uniformities in the device confinement potential, which makes the observation of ballistic transport challenging in longer wires,[10] and most significantly for our work, (b) it limits the mobility at low carrier densities, where the carrier-carrier interaction effects are strongest.[11] While it is possible to reduce the effect of remote ionized dopant scattering by increasing the separation between the ionized dopants and the 2D carrier layer, this also limits the minimum size of devices that can be defined. Hence, the goal of this present work is to develop mesoscopic hole devices where the modulation doping is absent.

In a standard modulation-doped device, the doping ensures that the device conducts at zero gate bias, and patterned surface gates are used to electrostatically deplete the carriers underneath it. One way to minimize remote ionized dopant scattering is to remove the modulation doping, in which case the device is non-conducting at zero gate-bias, and bias the gate to electrostatically accumulate carriers in the device and 'induce' conduction beneath the gate.[11-14] 1D electron quantum wires fabricated using an undoped AlGaAs/GaAs field effect transistors (FET) have shown conductance quantization[15] and have been used to study the density dependence and d.c. bias dependence of the 0.7-structure.[16-18] While that study was limited to interaction parameter $r_s \sim 1$, the development of induced p-type quantum wires will allow such experiments to be performed using holes, enabling studies of the 0.7-structure in the strongly interacting regime ($r_s > 10$).[19] Such studies will be a significant advance in understanding the role that many-body effects play in generating the 0.7-structure in 1D systems.

Here we report the development of a 1D hole quantum wire fabricated using an AlGaAs/GaAs FET structure without modulation doping. Our device shows clear quantized conductance plateaus at milli-Kelvin temperatures, including the 0.7-structure. We show that this device has very stable gate characteristics, which will enable studies at low densities where large values of $r_s$ and high mobilities can be achieved simultaneously. Finally, to show that these devices could be used to study 1D systems in the strong interaction regime, we present initial measurements of the density and temperature dependence of the device conductance.

Figure 1(a) shows a schematic cross-section of the AlGaAs/GaAs heterostructure used in this experiment. The active region is grown on a (311)A GaAs substrate and consists of a 1.5 μm GaAs layer, a 175 nm AlGaAs layer and a 25 nm GaAs layer, all of which are undoped. The final layer in the structure is a 75 nm heavily doped ($5 \times 10^{18}$ cm$^{-3}$ Si) p$^+$-GaAs layer, which is used as a metallic gate. When this gate is negatively biased ($\approx$ -0.1 V), a two-dimensional hole system (2DHS) forms at the AlGaAs/GaAs interface. The use of a p$^+$-gate,[14] as opposed to an n$^+$-gate,[12] allows the accumulation of holes at a lower gate bias due to the Fermi level pinning, thereby reducing current leakage between the gate and source/drain contacts. Electrical contact to the 2DHS was achieved using annealed AuBe ohmic contacts, which were deposited using a self-aligned process.[12] Electron beam lithography and wet etching (to a depth of 75 nm) were used to create the quantum wire by dividing the p$^+$-layer is into three separate, independently biasable gates – two "side-gates" and a "top-gate" (shown in the SEM micrograph in Fig. 1(b)).

All electrical measurements were performed using standard low-frequency lock-in techniques in a constant-voltage two-terminal configuration with an applied bias of 200 μV (most of this is dropped across the ohmic contacts, with $\approx$ 20 μV dropped across the 1D channel). Care was taken to ensure that the gate leakage current remained less than 0.1 nA throughout the experiment. The experiment was performed using a Kelvinox-100 dilution refrigerator with a base temperature of ~20 mK.

We now discuss in detail how our device operates. Applying sufficient negative bias $V_{TG}$ = -0.1 V to the top-gate establishes a 2DHS in areas directly underneath the top-gate (i.e., two large 2DHS reservoirs connected by a narrow quasi-1D channel). The hole density $p$ is determined by measuring the Hall slope in the 2DHS region adjacent to the wire, and controlled by tuning $V_{TG}$. The top gate operates over the bias range from the threshold voltage of $V_{TG}$ = -0.16 V to $V_{TG}$ = -0.41 V where the top gate starts to leak to the ohmic contacts (current leakage > 1 nA).

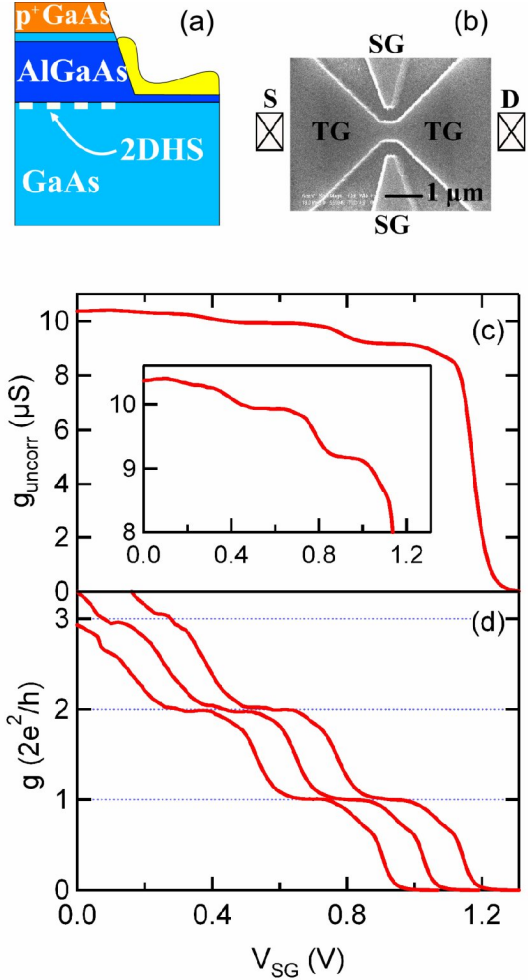

FIG. 1: (a) Schematic diagram of the heterostructure used in this study, (b) SEM micrograph of the 1D wire. S and D denote source and drain ohmic contacts, SG and TG stand for 'side gate' and 'top gate'; (c) conductance $g_{uncorr}$ of the 1D wire vs side gate voltage $V_{SG}$ prior to subtraction of the contact resistance $R_c$. The inset shows the same data magnified from 8 to 10.6 μS to highlight the presence of plateaus, and (d) contact-corrected conductance $g$ vs $V_{SG}$ (to demonstrate the device stability we show three consecutive sweeps for the same $V_{TG}$; traces are offset horizontally by -0.12 V for clarity).

We begin by setting a fixed top-gate bias of $V_{TG}$ = -0.36 V, which results in a hole density of $p = 1.18 \times 10^{11}$ cm$^{-2}$, (corresponding to $r_s$ = 11.5),[19] and giving a hole mobility of $7.1 \times 10^5$ cm$^2$/Vs. Applying a positive bias $V_{SG}$ to the side gates causes the 1D channel to narrow, reducing the conductance as shown in Fig. 1(c). The conductance ultimately drops to zero (i.e., the 1D channel 'pinches off') at $V_{SG} \sim 1.2$ V. Conductance plateaus are evident in the data in Fig. 1(c), confirming that the device operates as a 1D ballistic hole system. Note that at $V_{SG} = 0$ V in Fig. 1(c), the conductance is ~10 μS, much less than $2e^2/h$ (77 μS). This low conductance is due to the high resistance $R_c$ of the ohmic contacts adding in series with the quantum wire in the measurements. We find that $R_c$ is strongly dependant on $V_{TG}$ (ranging from $R_c \sim 80$ kΩ at $V_{TG} = -0.28$ V to $R_c \sim 150$ kΩ at $V_{TG} = -0.40$ V). Additionally, the *I-V* characteristics are slightly nonlinear with a maximal change in $R_c$ of 3.5 % per nA. Hence, in Fig. 1(d) and thereafter, we plot the conductance *g* after the subtraction of $R_c$. This contact-corrected conductance data shows plateaus well quantized in units of $2e^2/h$, as can be observed in Fig. 1(d). Most significantly and in contrast to previous studies,[4-7] the lower plateaus are clear and well defined, and we observe a strong and reproducible plateau at $g = 0.7 \times 2e^2/h$.

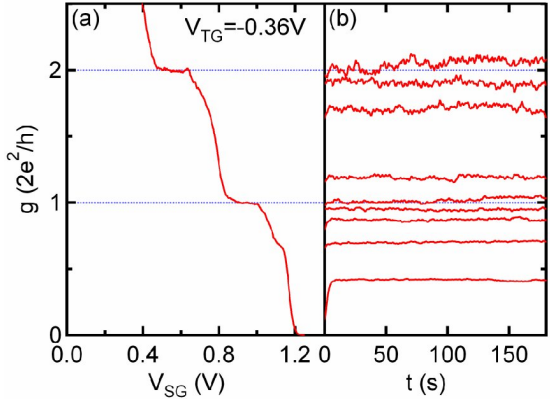

FIG. 2: (a) Conductance *g* vs $V_{SG}$ at $V_{TG} = -0.36$ V, and (b) *g* vs. time *t* for several set values of $V_{SG}$.

An on-going challenge in the development of p-type quantum wires has been instabilities in the electrical characteristics of these devices.[4-7] This typically results in noisy, poorly quantized conductance plateaus, and for a fixed gate bias configuration, large fluctuations in the device conductance over longer time scales (of order minutes to hours). To demonstrate the reproducibility and stability of our device, we firstly show three consecutively obtained traces in Fig. 1(d). Note that the pinch-off voltage for each of these traces is identical within 0.01 V, the side-gate voltage step size. We also show the time evolution of the conductance at various fixed $V_{SG}$ values at $V_{TG} = -0.36$ V in Fig. 2. At lower conductance, the device is remarkably stable with less than $0.02 \times 2e^2/h$ average conductance drift over several minutes. Even though the traces in Fig. 2(b) look more 'noisy' at higher conductance ($0.15 \times 2e^2/h$ average noise), the actual noise is in fact the same. The apparent increase in noise at large conductances is an artifact due to the 'stretching' of the higher plateaus that occurs with the subtraction of a large $R_c$, and is not due to temporal instabilities in the device. The stability of our device is in marked contrast to previous studies of modulation doped 1D hole systems[4,5,9] and is likely due to the absence of remote doping which introduces charge traps with long recombination times.

To study interaction effects, it is necessary to maximize $r_s$, but in electron systems it is hard to achieve $r_s \gg 1$ even at very low densities (~$10^{10}$ cm$^{-2}$). On the contrary, the larger hole mass means that $r_s$ exceeds 10 even at relatively high densities (~$10^{11}$ cm$^{-2}$). In Fig. 3 we plot the conductance *g* as a function of $V_{SG}$, for different hole densities *p* between $8.33 \times 10^{10}$ and $1.35 \times 10^{11}$ cm$^{-2}$. The bias $V_{SG}$ at which the quantum wire pinches off, as well as the strength and number of plateaus decreases with decreasing density. The higher plateaus appear to have weak resonances, but these are likely due to noise amplified at higher conductances (c.f. Fig. 2). The magnitude of the resonances is at most 4 pA out of a total source-drain current of 1.7 nA Therefore it is not possible to distinguish between the noise and resonances at higher conductances.

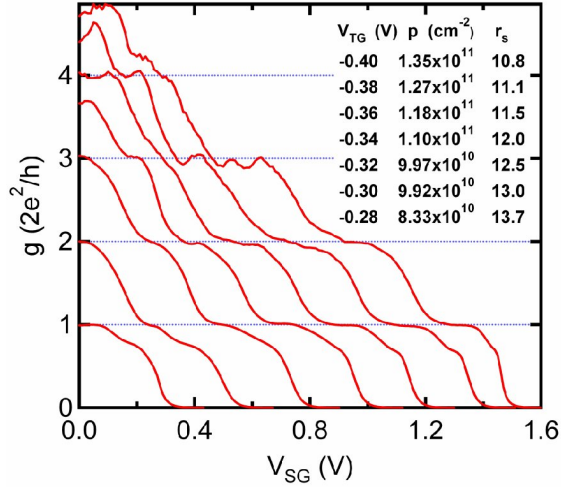

FIG. 3: Conductance $g$ vs. $V_{SG}$ at different values of the top gate voltage $V_{TG}$. $V_{TG}$ values and their corresponding $p$ and $r_s$ values are indicated in the legend.

Moreover, resonances inside the quantum wire would be strongest at low conductances due to reduced screening whereas our data shows the opposite trend. The number of plateaus observed in the data in Fig. 3 (up to five plateaus depending on $p$) is consistent with the 400 nm lithographic width of the 1D channel and a Fermi wavelength of ~ 70nm. We note that the leftmost curve in Fig. 3, obtained at $p = 8.3 \times 10^{10}$ cm$^{-2}$, shows only one plateau, and the 1D wire is pinched off at lower densities. This pinch-off density could be lowered by increasing the lithographic width of the wire, as the 2DHS remains measurable down to $2.5 \times 10^{10}$ cm$^{-2}$ (i.e., $r_s = 25$). A noteworthy feature of the data in Fig. 3 is that the 0.7-structure appears to evolve as a function of the hole density. An in-depth study of this effect will be reported elsewhere.

Finally, in Fig. 4 we explore the temperature $T$ dependence of the quantum wire conductance. The conductance traces in Fig. 4 have been offset horizontally for clarity, with low $T$ on the left to high $T$ on the right. The plateaus become thermally smeared with increasing temperature, with the $2e^2/h$ plateau disappearing at $T \sim 600$ mK, while higher plateaus weaken at lower $T$. This can be compared with electron systems,[20] where plateaus typically persist to higher temperatures of order ~ 4 K.

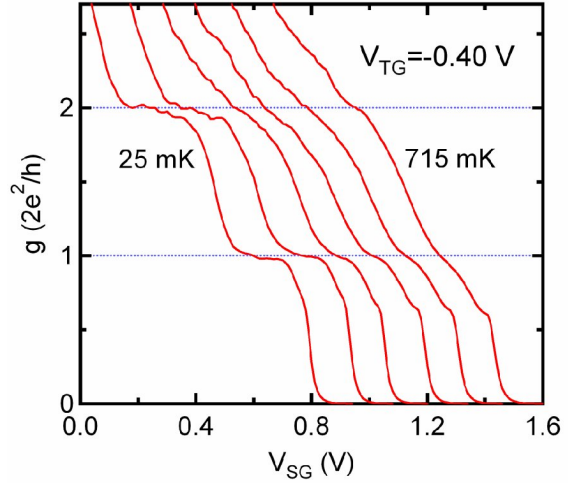

FIG. 4. Conductance of the quantum wire $g$ vs $V_{SG}$ for different temperatures $T = 25, 175, 370, 515, 620, 715$ mK. For clarity, the traces are offset sequentially by -0.12 V from right (715 mK, no offset) to left (25 mK, offset by –0.60 V).

The temperature dependence in Fig. 3 is also consistent with our source-drain bias measurements (not shown) which give a separation between the first two sub-bands of $\Delta E \approx 300$ μeV, much smaller than that found in electrons ($\Delta E \sim 1$ meV), due to the small mass of electrons compared to holes. Note that the plateau at $\sim 0.7 \times 2e^2/h$ becomes significantly stronger with increasing temperature. This is also consistent with the behavior reported in 1D electron systems,[3] suggesting that the 0.7 structure observed here has a similar origin to that observed in electron quantum wires.

In conclusion, we present results obtained from a 1D quantum wire fabricated using an AlGaAs/GaAs FET structure without modulation doping. Our 1D hole device exhibits very stable, clear conductance quantization, including the 0.7-structure. This device stability will enable studies at very low densities where strong interactions ($r_s > 10$) and high mobility can be achieved simultaneously, and thereby provide opportunities for studying the effects of spin-orbit coupling and strong carrier-carrier interactions in mesoscopic hole systems.

This work was funded by Australian Research Council (ARC). O.K. acknowledges support from the UNSW IPRS scheme. A.P.M. acknowledges an ARC Postdoctoral Fellowship.